\documentclass[11pt]{article}

\usepackage{amssymb,amsopn}

\newlength{\vshift}
\newlength{\hshift}
\def\la{\lambda}

\def\ds{\stackrel{\star}{,}}
\def\x{\hat x}

\def\p{\partial}

\def\lb{\lbrack}
\def\rb{\rbrack}

\begin{document}

\begin{titlepage}

$\,$

\vspace{1.5cm}
\begin{center}

{\LARGE{\bf A twisted look on kappa-Minkowski:\\
$U(1)$ gauge
theory}}

\vspace*{1.3cm}

{{\bf Marija Dimitrijevi\' c$^1$ and Larisa Jonke$^2$}}

\vspace*{1cm}

$^1$University of Belgrade, Faculty of Physics\\
Studentski trg 12, 11000 Beograd, Serbia \\[1em]

$^2$Theoretical Physics Division, Rudjer Bo\v skovi\' c Institute\\
Bijeni\v cka 54, 10000 Zagreb, Croatia\\

\end{center}

\vspace*{2cm}

\begin{abstract}
Kappa-Minkowski space-time is an example of noncommutative 
space-time with potentially interesting phenomenological consequences. However, the construction of field theories on this space, although operationally well-defined, is plagued  with ambiguities. A part of ambiguities can be resolved by clarifying the  geometrical picture of gauge transformations on the $\kappa$-Minkowski space-time. To this end we use the twist approach to construct the noncommutative $U(1)$ gauge theory coupled to fermions. 
However, in this approach we cannot maintain the kappa-Poincar\'e symmetry; the corresponding symmetry of the twisted kappa-Minkowski space is the
twisted igl(1,3) symmetry. We construct an action for the gauge and matter fields in a geometric way, as an integral of a maximal form. We use 
the Seiberg-Witten map to relate noncommutative and commutative degrees of freedom and expand the action to obtain the first order corrections in the deformation parameter.
\end{abstract}
\vspace*{1cm}



\vspace*{1cm}
\quad\scriptsize{eMail:
dmarija@ipb.ac.rs, larisa@irb.hr}
\vfill

\end{titlepage}\vskip.2cm

\newpage
\setcounter{page}{1}
\newcommand{\Section}[1]{\setcounter{equation}{0}\section{#1}}
\renewcommand{\theequation}{\arabic{section}.\arabic{equation}}

\section{Introduction}

It is generally believed that the picture of
space-time as a differentiable manifold should
break down at very short distances of the order
of the Planck length.  There are different
proposals for the modified space-time structure
which should provide consistent framework
encompassing physics in this regime.  These
proposals include, among others, the dynamical
triangulation as a way of direct
geometrical construction of modified space-time, strings
and loops as non-local fundamental observables dynamically
generating space-time, and a deformation of algebra of functions on a manifold
as a way of introducing a 'noncommutative space-time'.

The theoretical motivation for introducing a non-trivial algebra of
coordinates in order to modify the space-time structure comes from
various ideas and results. Historically, the first proposal by Snyder
\cite{Snyder} was put forward as a way of introducing a cut-off in
quantum field theory, i.e., as a proposal for regularization of
divergences. The quantum group approach \cite{QGbook} appeared as a
generalisation of concept of symmetries that should encompass physics
on a quantum manifold. Today, this idea is realised in the framework
of spin foam models based on the representation theory of quantum
groups \cite{nesto}. More recently, the realisation that the open string
theories and D-branes in the presence of a background
antisymmetric B-field give rise to noncommutative effective field
theories \cite{stringyNC, SW}  gave boost to research of  field
theories on a noncommutative space-time. It is believed that  the field theories on noncommutative spaces taken as effective models are capable of capturing some generic features of an elusive quantum theory of gravity.

The main advantage of such effective models is that one can
extract phenomenological
consequences of the space-time modification using the standard
field-theoretical tools. However, one needs to have a full
understanding of the symmetry structure of these
models and their renormalization properties to
be able to give  testable predictions.

In this work our primary interest is to examine
compatibility of the local gauge principle with the
deformation of algebra of functions on
a specific example of noncommutative space-time, the $\kappa$-Minkowski
space-time. The commutation relations of coordinates of 
the $m$ dimensional $\kappa$-Minkowski space-time
are of the Lie-algebra type
\begin{equation}
[\hat{x}^0, \hat{x}^j] = \frac{i}{\kappa}\hat{x}^j, \quad
[\hat{x}^i, \hat{x}^j] = 0, \label{IntroKom}
\end{equation}
where $i,j=1,\dots m-1$ and the zeroth component corresponds to the time direction.
One of the interesting properties of this noncommutative space-time is
that there is a quantum group symmetry acting on it. It is a
dimensionfull deformation of the global Poincar\' e group, the
$\kappa$-Poincar\' e group. The constant $\kappa$ has dimension of
energy and sets a deformation scale. Historically, the
$\kappa$-Poincar\' e group
was first obtained by Lukierski et al. in \cite{luk}
by the In\"{o}n\"{u}-Wigner contraction of
the $q$-anti-de Sitter Hopf algebra $SO_q(3,2)$. The $\kappa$-Poincar\' e
Hopf algebra was introduced in \cite{luk3} as a dual symmetry structure to
the $\kappa$-Poincar\' e group.  The $\kappa$-Minkowski space-time
is a module of this algebra. The $\kappa$-Poincar\' e group found its realisation in
the Doubly Special Relativity (DSR) theories
\cite{DSR}. These theories are introduced as a possible
generalisation of Special Relativity with one additional invariant
scale, usually taken to be an energy scale (of order of Planck energy)
or a length scale (of order of Planck length), see discussion in
\cite{RelLoc2}. The generalisation is done
is such a way that it leads to a
$\kappa$-Poincar\' e invariant modified dispersion relation for photons
and the energy-dependent speed of light. However, the claim of
\cite{Sabine} that this modification of the speed of light is 23 orders
of magnitude stronger than the recent measurements of gamma-ray bursts,
opened an intensive discussion, see \cite{Camelia-Sabine}. One of the
new ideas that originated form that discussion is the relativity of
locality \cite{RelLoc1}, an idea which should be relevant
in the regime characterised by negligible $\hbar$ and $G$
(classical-non gravitational regime), so that both
quantum and gravitational effects are small, with their ratio kept fixed.
In this new approach, the modified dispersion
relation is no longer invariant under the $\kappa$-Poincar\' e
action, but it transforms when the reference frame is changed. This
apparently leads to the modification of the speed of light that is of
the same order as the expected measured corrections and therefore
these two can be compared \cite{ExpDispRel}.

As we can see, the $\kappa$-Minkowski space-time is an example of
noncommutative space-time with potentially interesting phenomenological consequences. The construction of field theories on this space, although operationally well-defined, is plagued  with ambiguities. In our previous work \cite{miU1} we showed that within the framework defined in \cite{kappaft, kappagft} one
can consistently describe a gauge theory on the $\kappa$-Minkowski space-time
by explicit construction of $U(1)$ gauge theory coupled to fermions.
Although successful, our construction revealed certain ambiguities which
were fixed by the physical arguments and
intuition, rather then by the formalism itself. We lacked a better
understanding of the symmetries in the model and the geometrical formulation
of gauge theory, with the gauge field viewed as the connection 1-form.
With this motivation in mind, in this paper
we use the twist formalism in order to gain a better understanding of
the gauge theory on the $\kappa$-Minkowski space-time.

It is important to note that the twisted symmetry does not have 
the usual dynamical significance and there is no Noether procedure associated with it. In this paper we view this symmetry as a way of bookkeeping, a prescription that allow us to consistently apply deformation in the theory. The effective model obtained by expansion in the deformation parameter is  however amenable to the usual analysis, and is the one from  which one should draw out physical consequences of the deformation introduced. 

In order for the paper to be self-consistent,
in the next section we review some known results about the twisted differential
geometry. In Section 3 we construct the $\kappa$-Minkowski
space-time by choosing an explicit twist. Especially, we discuss  the differential structure on the obtained space-time: differential calculus, $\star$-algebra of
forms and integral. Using the mathematical tools introduced in
the previous sections,
in Section 4 the noncommutative $U(1)$ gauge theory coupled to fermions is constructed.
We use the Seiberg-Witten (SW) map to relate noncommutative and commutative degrees
of freedom. The action for the gauge and matter fields is written in a
geometric way, as an integral of a 
maximal form. We then expand the action up to first order in the 
deformation parameter, obtain equations of motions and discuss  
possible deformations of dispersion relations for free fields.  Finally, in Section 5
we discuss the obtained results and list some open questions and problems.

\section{Noncommutative spaces from a twist}

There are different ways to realize a noncommutative space and to formulate
physical models on it, see \cite{NCbooks} and \cite{NCbookMi}. One of
the most discussed
approaches is that of deformation quantization. In this approach a
noncommutative space is a quotient of the algebra freely generated
by the operators $\hat{x}^\mu$ and divided by the ideal generated by
the commutation relations
\begin{equation}
\lbrack \hat{x}^\mu,\hat{x}^\nu \rbrack = i\Theta^{\mu\nu}(\hat{x})
\label{1def.1}
\end{equation}
where $\Theta^{\mu\nu}(\hat{x})$ is an arbitrary polynomial of
$\hat{x}^\mu$ operators. This algebra can be represented on the space of
commuting coordinates for the most interesting and/or the most studied examples\footnote{The algebras need to  fulfil the
Poincar\' e-Birkoff-Witt property \cite{mssw}, and this is true for the canonical deformation, the Lie-algebra type of deformation  and the quantum group type of deformation. }.  The (noncommutative) 
algebra
multiplication between two functions of noncommuting coordinates
$\hat{f}$ and $\hat{g}$ is then mapped to the  $\star$-product:
\begin{eqnarray}
\hat{f}\cdot\hspace*{0.5mm}{\hat{g}}({\hat{x}})\mapsto
{f\star g(x) \in {\cal{A}}_x} .
\label{Intro-star}
\end{eqnarray}
Here $f$ and $g$ are functions of the commuting coordinates and
with ${\cal A}_x$ we label the algebra of functions on the commutative
space.
The product (\ref{Intro-star}) is bilinear and associative but noncommutative. The algebra
of noncommuting coordinates $\hat{\cal{A}}_{\hat{x}}$ is then
isomorphic to the algebra of commuting coordinates with the
$\star$-product (instead of the usual point-wise multiplication) as
a multiplication. A well known example is the
$\star$-product for the canonically deformed space defined by
\begin{equation}
\lbrack \hat{x}^\mu,\hat{x}^\nu \rbrack = i\theta^{\mu\nu} ,
\label{1def.2}
\end{equation}
where $\theta^{\mu\nu}$ is an antisymmetric constant
matrix of mass dimension minus two.
The $\star$-product is given by the Moyal-Weyl product
\begin{eqnarray}
&&f\star g\,(x) =
\lim _{x\to y}e^{\frac{i}{2}\theta^{\rho\sigma}\frac{\partial}
{\partial x^\rho}
\frac{\partial}{\partial y^\sigma}}
f(x)g(y) \label{mw}\\
&&= f\cdot g + \sum _{n=1}^\infty \Big(\frac{i}{2}\Big)^n
\frac{1}{n!}\theta^{\rho_1 \sigma_1}\cdots
\theta^{\rho_n \sigma_n}\Big(\partial_{\rho_1}\dots\partial_{\rho_n}f(x)
\Big)
\Big(\partial_{\sigma_1}\dots\partial_{\sigma_n}g(x)
\Big)
.\nonumber
\end{eqnarray}
Now one can define a noncommutative space as the usual space of
commuting coordinates with the point-wise multiplication replaced by a
noncommutative $\star$-product. Different models were constructed using
this approach. A noncommutative extension of the
Standard Model was constructed in \cite{NCSM} and some phenomenological
consequences were analyzed in \cite{NCSMCons}. Renormalization of
different models was discussed in \cite{NCRenorm}.

However, there is a drawback of this approach. Namely, it is not clear what
happens with symmetries of the theory in this approach. For example,
the commutation relations (\ref{1def.2}) obviously break the global Lorentz
symmetry, since $\theta^{\mu\nu}$ is constant. Is there a deformed
symmetry which replaces the global Lorentz symmetry in this case?
If it exists, what is it? An answer to these question
could be given using the twist formalism.

\subsection{Deformation by a twist}

\noindent The main idea of the twist formalism is to first deform the symmetry of the
theory and then see the consequences this deformation has on the space-time
itself. There is a well defined way to deform the symmetry Hopf algebra. In his
paper \cite{Drinfeld} Drinfel'd introduced a notion of twist. The
twist ${\cal F}$ is an invertible operator which belongs to
$Ug \otimes Ug$, where $Ug$ is the
universal enveloping algebra of the symmetry Lie algebra $g$. The universal
enveloping algebra $Ug$ is a Hopf algebra
\begin{eqnarray}
[t^a, t^b] &=& if^{abc}t^c, \nonumber\\
\Delta(t^a) &=& t^a \otimes 1 + 1\otimes t^a, \nonumber\\
\varepsilon (t^a) &=& 0,\quad S(t^a) = -t^a .\label{Ug}
\end{eqnarray}
In the first line $t^a$ label the generators of the symmetry algebra $g$ and
the structure constants are labelled by $f^{abc}$. In the second line the coproduct of the generator $t^a$ is given. It
encodes the Leibniz rule and specifies how the symmetry transformation acts on products
of fields/representations. In the last line, the counit and the
antipode are given. The properties which the twist $\cal{F}$ has
to satisfy are:
\begin{enumerate}

\item the cocycle condition
\begin{equation}
\label{propF11}
({\cal F}\otimes 1)(\Delta\otimes id){\cal F}=(1\otimes
{\cal F})(id\otimes \Delta){\cal F}, \label{Twcond1}
\end{equation}
\item normalization
\begin{equation}
(id\otimes \epsilon){\cal F} = (\epsilon\otimes id){\cal F}=1\otimes 1,
\label{Twcond2}
\end{equation}
\item perturbative expansion
\begin{equation}
{\cal F} = 1\otimes 1 + {\cal O}(\lambda), \label{Twcond3}
\end{equation}
\end{enumerate}
where $\lambda$ is a small deformation parameter. The last property is
not necessary. It provides an expansion around the undeformed case in the limit
 $\lambda\to 0$. We shall frequently use the notation (sum over $\alpha=1,2,...\infty $
is understood)
\begin{equation}
\label{Fff}
{\cal F} = {\rm f}^\alpha \otimes{\rm f}_\alpha, \quad
{\cal F}^{-1}=\bar{\rm f}^\alpha \otimes\bar{\rm f}_\alpha ,
\end{equation}
where, for each value of $\alpha$, $\bar{\rm f}^\alpha$ and
$\bar{\rm f}_\alpha$ are two distinct elements of $Ug$ (and similarly
${\rm f}^\alpha$ and ${\rm f}_\alpha$ are in $Ug$).

We also introduce the universal $\cal{R}$-matrix
\begin{equation}
{\cal R} = {\cal F}_{21}{\cal F}^{-1} ,\label{defUR}
\end{equation}
where by definition ${\cal F}_{21}={\rm f}_{\alpha}\otimes
{\rm f}^\alpha$.
In the sequel we use the notation
\begin{equation}
{\cal R} = R^\alpha\otimes R_\alpha ,\quad
{\cal R}^{-1} = \bar{R}^\alpha\otimes\bar{R}_\alpha.
\end{equation}

\subsection{Consequences of the twist}

\noindent The twist acts on the symmetry Hopf algebra and gives the twisted
symmetry Hopf algebra
\begin{eqnarray}
[t^a, t^b] &=& if^{abc}t^c,\nonumber\\
\Delta_{\cal F}(t^a) &=& {\cal F}\Delta (t^a){\cal F}^{-1} \nonumber\\
\varepsilon(t^a) &=& 0, \quad
S_{\cal F}(t^a) = {\rm f}^\alpha S({\rm f}_\alpha)S(t^a)S(\bar{\rm f}^\beta)
\bar{\rm f}_\beta. \label{TwistedUg}
\end{eqnarray}
We see that the algebra remains the same, while in general the
comultiplication changes. This leads to the deformed Leibniz rule for the
symmetry transformations when acting on product of fields.

We can now use the twist to
deform the commutative geometry on space-time (vector fields, 1-forms,
exterior algebra of forms, tensor algebra).
The guiding principle is the observation that every time we have
a bilinear map
$$\mu\,: X\times Y\rightarrow Z ,$$
where $X,Y,Z$ are vector spaces
and when there is an action of the Lie algebra $g$
(and therefore of ${\cal F}^{-1}$) on $X$ and $Y$
we can combine the map $\mu$ with the action of the twist. In this way
we obtain the deformed map $\mu_\star$:
\begin{equation}
\mu_\star = \mu {\cal F}^{-1}. \label{mustar}
\end{equation}
The cocycle condition (\ref{propF11}) implies that if $\mu$ is an
associative product then also $\mu_\star$ is an associative product.

Let us analyze this deformation in more detail. For convenience we 
now consider one particular class of twists, the Abelian twists
\begin{equation}
{\cal F} = e^{-\frac{i}{2}\theta^{ab}X_a\otimes X_b} .\label{AbTwist} 
\end{equation}
Here $\theta^{ab}$ is a constant antisymmetric matrix, $a,b=1,2, \dots p\leq m$
and $X_a=X_a^\mu\partial_\mu$ are commuting vector fields, $[X_a,X_b] =0$.
The algebra of vector fields on the space-time $M$ we label with $\Xi$
and the universal enveloping algebra of this algebra with $U\Xi$. 
Then ${\cal F}$ belongs to $U\Xi \otimes U\Xi$. In the view of 
(\ref{Ug})-(\ref{Twcond3}), the symmetry algebra is the algebra of
diffeomorphisms generated by vector fields $\xi=\xi^\mu\partial_\mu\in \Xi$. 
Note that depending on the choice of vector fields $X_a$ one can also
consider a subalgebra of the diffeomorphism algebra such as Poincar\' e
or conformal algebra.

Applying the inverse of the twist (\ref{AbTwist}) to the usual 
point-wise multiplication of functions on the space-time $M$, 
$\mu(f\otimes g)=f\cdot g$, we obtain the $\star$-product of
functions
\begin{eqnarray}
f\star g &=& \mu {\cal F}^{-1}(f\otimes g)\nonumber\\
&=& \bar{\rm f}^\alpha(f) \bar{\rm f}_\alpha(g) \nonumber\\
&=& \bar{R}^\alpha(g)\star\bar{R}_\alpha(f). \label{FunctionsStar}
\end{eqnarray}
We see that the $R$-matrix encodes the
noncommutativity of the $\star$-product.
The action of the twist ($\bar{{\rm f}}^\alpha$ and $\bar{{\rm f}}_\alpha$) on the functions $f$ and $g$ is via the Lie derivative.

The product between functions and 1-forms is given again by following
the general prescription
\begin{equation}
h\star\omega = \bar{\rm f}^\alpha(h)\bar{\rm f}_\alpha(\omega)
\end{equation}
with an arbitrary 1-form $\omega$. The action of $\bar{{\rm f}}_\alpha$ on
forms is (again) given via the Lie derivative.
Functions can be multiplied from the left or from the right,
\begin{equation}
h\star\omega = \bar{\rm{f}}^\alpha(h)\bar{\rm{f}}_\alpha(\omega)
={\bar{R}^\alpha}(\omega)\star \bar{R}_{\alpha}(h).
\label{FunFormStar}
\end{equation}

Exterior forms form an algebra with the wedge product
$\wedge :\,\Omega^{\mbox{\boldmath $\cdot$}}\times
\Omega^{\mbox{\boldmath $\cdot$}}\rightarrow \Omega^{\mbox{\boldmath $\cdot$}}$.
We $\star$-deform the wedge product on two arbitrary forms $\omega$
and $\omega'$ into the $\star$-wedge product,
\begin{equation}
\omega\wedge_\star\omega' = \bar{\rm f}^\alpha(\omega)
\wedge \bar{\rm f}_\alpha(\omega').\label{WedgeStar}
\end{equation}
We denote by $\Omega^{\mbox{\boldmath $\cdot$}}_\star$
the linear space of forms equipped with the $\star$-wedge product
$\wedge_\star$.

As in the commutative case, the exterior forms are totally
$\star$-antisymmetric (contravariant) tensor-fields.
For example, the 2-form
$\omega\wedge_\star\omega'$ is the $\star$-antisymmetric combination
\begin{eqnarray}
\omega\wedge_\star\omega' &=& \bar{\rm f}^\alpha(\omega)
\wedge \bar{\rm f}_\alpha(\omega') \nonumber\\
&=&\omega\otimes_\star\omega'
-\bar{R}^\alpha(\omega')\otimes_\star \bar{R}_{\alpha}(\omega),
\label{Star1Forms}\\
&=& -\bar{R}^\alpha(\omega')\wedge_\star \bar{R}_{\alpha}(\omega) ,
\nonumber
\end{eqnarray}
with the $\star$-tensor product defined as
\begin{equation}
T_1\otimes_\star T_2 = \bar{\rm f}^\alpha(T_1)
\otimes \bar{\rm f}_\alpha(T_2). \label{TensPrStar}
\end{equation}

The usual exterior derivative
${\rm d}: {\cal A}_x \rightarrow \Omega$ satisfies the Leibniz rule\footnote{The reason for this is that the usual exterior derivative commutes with the Lie derivative.}
${\rm d} (f\star g) = {\rm d} f\star g + f\star {\rm d}g $ and is
therefore also the $\star$-exterior derivative. One can rewrite the
usual exterior derivative of a function using the $\star$-product as
\begin{eqnarray}
{\rm d}f &=& (\partial_\mu f){\rm d}x^\mu \nonumber\\
&=& (\partial^\star_\mu f)\star {\rm d}x^\mu, \label{ExtDeriv}
\end{eqnarray}
where the new derivatives $\partial_\mu^\star$ are defined by this
equation.

The usual integral is cyclic under the $\star$-exterior products of
forms, that is up to boundary terms we have
\begin{equation}
\int \omega_1 \wedge_\star \omega_2 = (-1)^{d_1\cdot d_2}
\int \omega_2 \wedge_\star \omega_1, \label{IntCycl}
\end{equation}
where $d=deg(\omega)$, $d_1+d_2 = m$ and $m$ is the dimension of the
space-time $M$. This property holds for the Abelian twist 
(\ref{AbTwist}). More generally, one can show \cite{PL1} that this 
property holds for any twist that satisfies the condition 
$S(\bar{\rm f}^\alpha)\bar{\rm f}_\alpha =1$,
with the antipode $S$.

\section{Kappa-Minkowski via twist}

Algebraically, the $m$-dimensional $\kappa$-Minkowski space-time can be
introduced as a  quotient
of the algebra freely generated by the coordinates $\hat{x}^\mu$ and 
divided by
the ideal generated by the following commutation relations:
\begin{equation}
\label{2.1}
\lb \x^\mu , \x^\nu \rb = i C^{\mu\nu}_\rho \x^\rho,
\quad \mu,\nu,\rho=0,\dots,m-1.
\end{equation}
Defining
\begin{equation}
\label{2.2}
 C^{\mu\nu}_\rho=a(\delta^\mu _0 \delta^\nu_\rho
-\delta^\nu_0\delta^\mu_\rho)
\end{equation}
the commutation relations (\ref{2.1}) can be rewritten as
\begin{equation}
[\hat{x}^0, \hat{x}^j] = ia\hat{x}^j, \quad
[\hat{x}^i, \hat{x}^j] = 0. \label{Intro2Kom}
\end{equation}
The metric of the $\kappa$-Minkowski space-time is
$\eta^{\mu\nu}=diag(1,-1,\ldots,-1)$.
The deformation parameter $a$ is related to the
frequently used parameter $\kappa$ as $a=1/\kappa$.
Latin indices denote space dimensions, zero the time
dimension and the Greek indices refer to all $m$ dimensions.

As we have said in the previous section there exists an isomorphism
between the abstract algebra and
the algebra of functions of commuting coordinates equipped with a
$\star$-product. There are different $\star$-product realizations
of the $\kappa$-Minkowski space-time, see \cite{StarPrKappa}.
The  symmetric $\star$-product for the $\kappa$-Minkowski
space-time, up to the first order in the deformation parameter, is given by
 \begin{eqnarray}
\label{sy3}
f(x)\star_{SO} g(x) &=&
f(x) g(x) + \frac{i}{2}C^{\mu\nu}_\lambda x^\lambda \p_\mu f(x) \p_\nu g(x) \nonumber\\
&=& f(x) g(x) + \frac{ia}{2}x^j \big( \p_0 f(x)\p_j g(x)
- \p_jf(x) \p_0  g(x)\big)  .
\end{eqnarray}
Using this $\star$-product, field theories on $\kappa$-Minkowski were
constructed \cite{FThKappaOni, kappaft}. However, there are some open problems in
this approach. We mention two of them. The ordinary partial derivative
$\partial_\mu=\frac{\partial}{\partial x^\mu}$ has a deformed Leibniz
rule due to the $x$-dependence of the $\star_{SO}$-product,
\begin{equation}
\partial_\mu (f\star_{SO} g) = (\partial_\mu f) \star_{SO} g 
+ f\star_{SO} (\partial_\mu g)
+ f (\partial_\mu \star_{SO}) g. \nonumber
\end{equation}
This property can lead to gauge fields and
field strength given in terms of higher order differential operators
\cite{miU1}. From the algebraic point of view there are different choices of derivatives
and one has to specify a criterion (e.g., transformation under
$\kappa$-Lorentz symmetry, convenient Leibniz rule) for choosing
one particular set\footnote{For different approaches to the problem of differential 
calculus on $\kappa$-Minkowski space-time see \cite{DiffCalcKappa}.}. 
The definition of an integral is also a problem. The
usual integral is not cyclic (again due to the $x$-dependence of the
$\star$-product) and one has to introduce the measure function $\mu$
in order to make it cyclic\footnote{Scalar field theory with
a cyclic integral and without a measure function was constructed and
discussed in \cite{TrampMelj}.},
\begin{equation}
\int {\rm d}^{m} x \mu(x) f\star_{SO} g
=  \int {\rm d}^{m} x \mu(x) g\star_{SO} f .\nonumber
\end{equation}
The equality holds up to boundary terms. In general, the measure function
spoils the symmetry properties of an action and of the corresponding equations of motion.
It also spoils the commutative limit since it is $a$-independent and does
not vanish in the limit $a\to 0$.  

In order to overcome some of these problems in this paper we follow the
twist approach. The choice of twist is not unique and it depends on the
properties that we want to obtain/preserve. We choose the following twist
\begin{eqnarray}
{\cal F} &=& e^{-\frac{i}{2}\theta^{ab}X_a\otimes X_b}\nonumber\\
&=& e^{-\frac{ia}{2}
(\partial_0\otimes x^j\partial_j-x^j\partial_j\otimes \partial_0)},
\label{KappaTwist}
\end{eqnarray}
with two commuting vector fields $X_1=\partial_0$ and $X_2=x^j\partial_j$
and
\begin{equation}
\theta^{ab} =\left( {\begin{array}{cc}
 0 &  a \\ -a &  0
\end{array} } \right)
. \nonumber
\end{equation}
This twist fulfils the conditions (\ref{Twcond1}), (\ref{Twcond2}) and
(\ref{Twcond3})
with the small deformation parameter $\lambda=a$. 
Deformed symmetry concerned, note that $X_2$ is not in
the universal enveloping algebra of the Poincar\' e algebra. Therefore we
have to enlarge the Poincar\' e algebra $iso(1,m-1)$ to the inhomogeneous general
linear algebra $igl(1,m-1)$ and twist this algebra instead of 
$iso(1,m-1)$. The generators (given in the representation on the
space of functions/fields) and the commutation relations
of $igl(1,m-1)$ are
\begin{eqnarray}
&& M_{\mu\nu} = x_\mu\partial_\nu , \quad P_\mu =\partial_\mu,\nonumber\\
&& \lbrack P_\mu, P_\nu \rbrack = 0, \quad
\lbrack M_{\mu\nu}, P_\rho\rbrack = \eta_{\mu\rho}P_\nu,\nonumber\\
&& \lbrack M_{\mu\nu}, M_{\rho\sigma}\rbrack = \eta_{\nu\rho}M_{\mu\sigma}
- \eta_{\mu\sigma}M_{\rho\nu} . \label{igl(1n)}
\end{eqnarray}
Let us discuss the consequences of the twist (\ref{KappaTwist}).

\subsection{Twisted symmetry}

\noindent The action of the twist (\ref{KappaTwist}) on the $igl(1,m-1)$
algebra follows from (\ref{TwistedUg}) and it has been analysed in detail
in \cite{Pachol}. Let us just summarise the most important results.
The algebra (\ref{igl(1n)}) remains the same.
On the other hand, since $X_2 = x^j\partial_j$ does not commute with
the generators $P_\mu$ and $M_{\mu\nu}$ the comultiplication
and the antipode change. Here we just give the result for the twisted
comultiplication, the other results can be found in \cite{Pachol}.
\begin{eqnarray}
\Delta P_0 &=& P_0 \otimes 1 + 1\otimes P_0, \nonumber\\
\Delta P_j &=& P_j \otimes e^{-\frac{i}{2}aP_0}
+ e^{\frac{i}{2}aP_0}\otimes P_j, \nonumber\\
\Delta M_{ij} &=& M_{ij} \otimes 1 + 1\otimes M_{ij}, \nonumber\\
\Delta M_{0j} &=& M_{0j} \otimes e^{-\frac{i}{2}aP_0}
+ e^{\frac{i}{2}aP_0}\otimes M_{0j} -\frac{i}{2}aP_j\otimes \mathrm{D}
+ \frac{i}{2}a\mathrm{D}\otimes P_j, \nonumber\\
\Delta M_{j0} &=& M_{j0} \otimes e^{-\frac{i}{2}aP_0}
+ e^{\frac{i}{2}aP_0}\otimes M_{j0} ,\nonumber\\
\Delta M_{00} &=& M_{00} \otimes 1 + 1\otimes M_{00}
-\frac{i}{2}aP_0\otimes \mathrm{D}
+ \frac{i}{2}a\mathrm{D}\otimes P_0. \label{TwistedCopr}
\end{eqnarray}
We introduced the notation $\mathrm{D}=x^j\partial_j$. Note that
$\kappa$-Poincar\' e symmetry
found in \cite{luk} will not be a symmetry of our twisted
$\kappa$-Minkowski space. The corresponding symmetry of the twisted
$\kappa$-Minkowski space is the twisted $igl(1,m-1)$ symmetry.

\subsection{$\star$-product}

\noindent The inverse of the twist (\ref{KappaTwist}) defines the $\star$-product
between functions/fields on the $\kappa$-Minkowski space-time
\begin{eqnarray}
f\star g &=& \mu_\star \{ f\otimes g \} \nonumber\\
&=& \mu \{ {\cal F}^{-1}\, f\otimes g\} \label{StarDef}\\
&=& \mu \{ e^{\frac{ia}{2}
(\partial_0\otimes x^j\partial_j-x^j\partial_j\otimes \partial_0)}
f\otimes g\}
\nonumber\\
&=& f\cdot g + \frac{ia}{2} x^j\big( (\partial_0 f) \partial_j g
-(\partial_j f) \partial_0 g\big) + {\cal O}(a^2) \nonumber\\
&=& f\cdot g + \frac{i}{2}C^{\rho\sigma}_\lambda x^\lambda
(\partial_\rho f)\cdot (\partial_\sigma g) + {\cal O}(a^2)
, \label{StarPrExp}
\end{eqnarray}
with $C^{\rho\sigma}_\lambda$ given in (\ref{2.2})
This product is associative, noncommutative and hermitean
\begin{equation}
\overline{f\star g} = \bar{g} \star \bar{f}. \nonumber
\end{equation}
The usual complex conjugation we label with ``bar''. In the zeroth order
(\ref{StarPrExp}) reduces to the usual point-wise multiplication. Note that
the first order term of this $\star$-product is the same as the first
order term of the symmetric $\star_{SO}$-product (\ref{sy3}). The second and higher
orders will be different. Of course, we obtain
\begin{equation}
[x^0 \ds x^j] = x^0\star x^j -
x^j \star x^0 = ia x^j,\quad [x^i \ds x^j] =0 .\label{xStarComm}
\end{equation}

\subsection{Twisted differential calculus}

\noindent One of the advantages of the twist formalism is the straightforward way
to define a differential calculus. Namely, as said in the previous section,
we just adopt the undeformed differential calculus with the following
properties
\begin{eqnarray}
{\rm d} (f\star g) &=& {\rm d}f\star g + f\star {\rm d}g,\nonumber\\
{\rm d}^2 &=& 0,\nonumber\\
{\rm d} f &=& (\partial_\mu f) {\rm d}x^\mu = (\partial^\star_\mu f)
\star {\rm d}x^\mu. \label{Differential}
\end{eqnarray}
The basis one forms are ${\rm d}x^\mu$. Knowing that the action of a
vector field on a form is given via Lie derivative one can show that
\begin{equation}
X_1 ({\rm d}x^\mu) =0, \quad
X_2 ({\rm d}x^\mu) = \delta ^\mu_j {\rm d}x^j.\label{LieDerdx}
\end{equation}
Using these relations one obtains that the basis 1-forms anticommute but do not
$\star$-commute with functions. They are not frame 1-forms in the sense
of Madore \cite{NCbooks}. Instead they fulfil
\begin{eqnarray}
{\rm d}x^\mu \wedge_\star {\rm d}x^\nu &=& {\rm d}x^\mu \wedge {\rm d}x^\nu
= - {\rm d}x^\nu \wedge {\rm d}x^\mu =  -{\rm d}x^\nu \wedge_\star
{\rm d}x^\mu,
\nonumber\\
f\star {\rm d}x^0 &=& {\rm d}x^0 \star f,\quad
f\star {\rm d}x^j = {\rm d}x^j \star e^{ia\partial_0}f. \label{fstardx}
\end{eqnarray}
Arbitrary 1-forms $\omega_1= \omega_{1\mu}\star {\rm d}x^\mu$ and
$\omega_2= \omega_{2\mu}\star {\rm d}x^\mu$ do not anticommute
\begin{equation}
\omega_1\wedge_\star \omega_2 =
- \bar{R}^\alpha(\omega_2)\wedge_\star \bar{R}_{\alpha}(\omega_1),
\label{anticom1forme}
\end{equation}
where the inverse of the ${\cal R}$ matrix  is given by
\begin{equation}
{\cal R}^{-1} = {\cal F}^2 = e^{-ia(\partial_0\otimes x^j\partial_j -
x^j\partial_j\otimes \partial_0)}. \label{Rbar}
\end{equation}
The $\star$-derivatives follow from (\ref{Differential}) and are given by
\begin{eqnarray}
&&\partial^\star_0 = \partial_0, \quad \partial^\star_j
= e^{-\frac{i}{2}a\partial_0}\partial_j,\nonumber\\
&&\partial^\star_0 (f\star g) = (\partial^\star_0 f)\star g +
f\star (\partial^\star_0 g), \nonumber\\
&& \partial^\star_j (f\star g) =
(\partial^\star_j f)\star e^{-ia\partial_0}g +
f\star (\partial^\star_j g). \label{ParcLeibniz}
\end{eqnarray}

\subsection{Integral}

\noindent The usual integral of a maximal form is cyclic
\begin{equation}
\int \omega_1 \wedge_\star \omega_2 = (-1)^{d_1\cdot d_2}
\int \omega_2 \wedge_\star \omega_1, \label{kappaIntCycl}
\end{equation}
with $d=deg(\omega)$ and  $d_1+d_2 = m$. Since basis 1-forms anticommute the volume
form remains undeformed
\begin{equation}
{\rm d}^{m}_\star x := {\rm d}x^0\wedge_\star{\rm d}x^1\wedge_\star
\dots {\rm d}x^{m-1} = {\rm d}x^0\wedge{\rm d}x^1\wedge\dots
{\rm d}x^{m-1} = {\rm d}^{m} x .\label{VolForm}
\end{equation}

\section{$U(1)$ gauge theory}

In this section we formulate a noncommutative $U(1)$ gauge theory coupled
to fermi\-on\-ic matter. We follow the Seiberg-Witten method \cite{SW} and the enveloping
algebra approach \cite{EnvAlg}. From now on we work in four dimensions. The
method is however general, it can be applied to any $SU(N)$ and $U(N)$
gauge group and in any number of dimensions.

The basic assumption of the SW map is that the noncommutative fields
and the noncommutative gauge parameter can be expressed
as functions of the commutative fields and the commutative gauge 
parameter $\alpha$. For example, the noncommutative gauge parameter
$\Lambda$ is
\begin{equation}
\Lambda = \Lambda(\alpha, A_\mu^0) := \Lambda_\alpha(A_\mu^0)  
\end{equation}
with
the commutative\footnote{In the following, superscript zero denotes undeformed, commutative fields.} gauge field $A_\mu^0$. The explicit form
of this dependence is found by solving the appropriate equations. In that way the number of degrees
of freedom in the noncommutative theory reduces to the number of degrees
of freedom of the corresponding commutative theory. 

\subsection{Matter fields}

The infinitesimal noncommutative gauge transformation of the field $\psi$
is given by
\begin{equation}
\delta^\star_\alpha \psi = i \Lambda_\alpha \star
\psi, \label{NCTrPsi}
\end{equation}
where $\Lambda_\alpha$ is the noncommutative gauge parameter related via
SW map with the commutative gauge parameter $\alpha$ and $\psi$
is the noncommutative matter field in the fundamental representation.
We demand consistency, that is that the algebra of gauge transformation closes:
\begin{equation}
(\delta^\star_\alpha \delta^\star_\beta -\delta^\star_\beta
\delta^\star_\alpha )\psi =  \delta^\star_{-i[\alpha,\beta]}\psi
.\label{swalg}
\end{equation}
In order to solve this equation, we expand $\Lambda_\alpha$ in the orders
of the deformation parameter
\begin{equation}
\Lambda_\alpha = \alpha + \Lambda_\alpha^1+ \dots + \Lambda_\alpha^k
+\dots .\nonumber
\end{equation}
Also we have to
expand the $\star$-product in the equation (\ref{swalg}). All the
expansions in this paper will be up to
first order in the deformation parameter $a$. The inhomogeneous equation
for $\Lambda_\alpha^1$ then reads
\begin{eqnarray}
&&\delta_\alpha\Lambda^1_\beta -\delta_\beta\Lambda^1_\alpha - i\lb\alpha,\Lambda^1_\beta\rb -i \lb\Lambda^1_\alpha,\beta\rb -\Lambda^1_{-i[ \alpha,\beta]}
= -\frac{1}{2}C^{\rho\sigma}_\lambda x^\lambda
\{\partial_\rho \alpha, \partial_\sigma\beta\},
\nonumber \\
&& \delta_\alpha\Lambda^1_\beta -\delta_\beta\Lambda^1_\alpha =
-C^{\rho\sigma}_\lambda x^\lambda
(\partial_\rho \alpha) (\partial_\sigma\beta)
.\label{Lam1}
\end{eqnarray}
In the second line we used the fact that in the case of $U(1)$ gauge
symmetry all commutators vanish and all anticommutators just add.
Note that $\delta_\alpha\Lambda^1_\beta \neq 0$ since $\Lambda^1_\beta$
is a function of the commutative gauge parameter $\beta$ and the commutative
gauge field $A_\mu^0$ and $\delta_\alpha A_\mu^0
= \partial_\mu \alpha \neq 0$. The solution of equation (\ref{Lam1}) is given by
\begin{equation}
\Lambda^1_{\alpha} = -\frac{1}{2}C^{\rho\sigma}_\lambda x^\lambda
A^0_\rho \partial_\sigma\alpha. \nonumber
\end{equation}
This solution is not unique, one can always add a solution of the
homogeneous equation to it. This is the freedom in the SW map. In the
case of $U(1)$ gauge group the only homogeneous term is of the form
\begin{equation}
\Lambda_\alpha^{\rm{hom}} = c_1C^{\rho\sigma}_\lambda x^\lambda
F^{0}_{\rho\sigma}\alpha ,\nonumber
\end{equation}
with the commutative field-strength tensor $F^{0}_{\rho\sigma}= 
\partial_\rho A_\sigma^0 - \partial_\sigma A^0_\rho$.
However, this term does not lead to a solvable equation for the noncommutative
gauge field and therefore we shall not consider it. The noncommutative gauge parameter up
to first order in the deformation parameter reads
\begin{equation}
\Lambda_{\alpha} = \alpha -\frac{1}{2}C^{\mu\nu}_\lambda x^\lambda
A^0_\mu \partial_\nu\alpha. \label{SWLambda}
\end{equation}

The solution for the matter field $\psi$ follows from (\ref{NCTrPsi})
and (\ref{SWLambda}) and is given by
\begin{equation}
\psi = \psi^0 -\frac{1}{2}C^{\rho\sigma}_\lambda x^\lambda
A^0_\rho(\partial _\sigma\psi^0) + id_1C^{\rho\sigma}_\lambda
x^\lambda F^0_{\rho\sigma}\psi ^0
+ d_2 aD^0_0\psi ^0 .\label{SWPsi}
\end{equation}
The terms with the real undetermined  coefficients $d_1$ and $d_2$ are the solutions of the homogeneous equation and represent the freedom of the SW map.

\subsection{Gauge fields}

\noindent In order to write a gauge invariant action for the matter field $\psi$
one has to introduce a covaraint derivative and a connection. We have
a preferred differential calculus on the $\kappa$-Minkowski space-time
given by (\ref{Differential}) and we will use it now. The covariant
derivative $D\psi$  is defined in the following way
\begin{eqnarray}
D\psi &=& d\psi-iA\star \psi \>=\> D_\mu^\star \psi \star {\rm d}x^\mu ,
\label{DPsi}\\
D_0^\star \psi= \partial_0^\star\psi \hspace*{-9mm}&&-\,iA_0\star\psi, \quad
D_j^\star \psi= \partial_j^\star\psi -iA_j\star e^{-ia\partial_0}\psi,
\end{eqnarray}
where the noncommutative connection $A = A_\mu \star {\rm d}x^\mu$
is introduced. The term $e^{-ia\partial_0}\psi$ comes from
$\star$-commuting $\psi$ through ${\rm d}x^j$. The transformation law
of the covaraint derivative
\begin{equation}
\delta_\alpha^\star D\psi = i\Lambda_\alpha\star D\psi \label{TrDPhi}
\end{equation}
defines the transformation law of the noncommutative connection. It is
given by
\begin{equation}
\delta_\alpha^\star A = {\rm d}\Lambda_\alpha + i[\Lambda_\alpha \ds A],
\label{ATr}
\end{equation}
or in the components
\begin{eqnarray}
\delta_\alpha^\star A_0 &=& \partial_0 \Lambda_\alpha
+ i [\Lambda_\alpha \ds A_0], \label{AnTr}\\
\delta^\star_\alpha A_j &=& \partial_j^\star \Lambda_\alpha
+ i\Lambda_\alpha\star A_j
- iA_j\star e^{-ia\partial_0}\Lambda_\alpha. \label{AjTr}
\end{eqnarray}
Assuming that $A_\mu = A_\mu^0 + A_\mu^1 + \dots$ one finds
the solutions of (\ref{AnTr}) and (\ref{AjTr})
\begin{eqnarray}
A_\mu &=& A^0_\mu - \frac{a}{2}\delta^j_{\mu}\Big(i\partial_0 A^0_j
+ A^0_0A^0_j\Big)+ {1\over 2} C^{\rho\sigma}_\lambda x^\lambda
\Big(  F^0_{\rho \mu}A^0_\sigma
- A^0_\rho \partial _\sigma A^0_\mu\Big) \nonumber\\
&& + d_3 C^{\rho\sigma}_\la x^\la \partial_\rho F^0_{\sigma\mu}
+ d_4 a F_{\mu 0}.\label{SWAmu}
\end{eqnarray}
The terms with the real, undetermined coefficients $d_3$ and $d_4$ are the solutions of the
homogeneous equation and represent the freedom of the SW map.
Note that the connection 1-form $A$ is real, but the components
$A_\mu$ are not necessarily real due to the $\star$-product in
$A = A_\mu\star {\rm d}x^\mu$.

In the next step we construct the field-strength tensor. The field-strength
tensor is a two-form given by\footnote{Equivalently, one can define the 
field-strength tensor as  $D^2\psi=-iF\star \psi$, with the covariant derivative $D$ given in (\ref{DPsi}).}
\begin{equation}
F = \frac{1}{2} F_{\mu\nu} \star {\rm d}x^\mu \wedge_\star {\rm d}x^\nu
= {\rm d}A - i A\wedge_\star A, \label{F}
\end{equation}
or in components
\begin{eqnarray}
F_{0j} &=& \partial_0^\star A_j - \partial_j^\star A_0
-iA_0\star A_j + iA_j\star e^{-ia\partial_0}A_0
,\label{Fnj}\\
F_{ij} &=& \partial_i^\star A_j - \partial_j^\star A_i
-iA_i\star e^{-ia\partial_0}A_j + iA_j\star e^{-ia\partial_0}A_i .
\label{Fij}
\end{eqnarray}
One can check that the field-strength tensor  transforms covariantly,
\begin{equation}
\delta_\alpha^\star F = i [\Lambda \ds F]. \label{FTr}
\end{equation}
Inserting the solution (\ref{SWAmu}) into (\ref{Fnj}) and (\ref{Fij})
results in
\begin{eqnarray}
F_{0j} &=& F_{0j}^0 -\frac{ia}{2}\partial_0F_{0 j}^0 -aA_0^0F_{0j}^0
+ C^{\rho\sigma}_\lambda x^\lambda\Big( F^0_{\rho 0}F_{\sigma j}^0
-A^0_\rho\partial _\sigma F_{0j}^0 \Big) \nonumber\\
&&  + a(d_3-d_4) \partial_0 F_{0j}^0
,\label{SWFnj}\\
F_{ij} &=& F_{ij}^0 -ia\partial_0 F_{ij}^0 - 2aA_0^0F_{ij}^0
+ C^{\rho\sigma}_\lambda x^\lambda \Big( F^0_{\rho i}F_{\sigma j}^0
- A^0_\rho\partial _\sigma F_{ij}^0 \Big)
\nonumber\\
&& + a(d_3-d_4)\partial_0F_{ij}^0 . \label{SWFij}
\end{eqnarray}

\subsection{Gauge field action}

\noindent In the commutative gauge theory one writes the action for the gauge
field using the Hodge dual of the field-strength tensor, $*F^0$:
\begin{eqnarray}
&&S_{g}^0 = \int F^0\wedge (*F^0), \nonumber\\
&&*F^0 = \frac{1}{2}\epsilon_{\mu\nu\alpha\beta}F^{0\alpha\beta}
{\rm d}x^\mu\wedge {\rm d}x^\nu.\nonumber
\end{eqnarray}
The indices on $F^{0\alpha\beta}$ are raised with the flat metric
$\eta_{\mu\nu}$ and
\begin{equation}
\delta_\alpha (*F^0) = i [\alpha, *F^0] = 0 \nonumber
\end{equation}
since we work with $U(1)$ gauge theory.

We try to generalise this to the $\kappa$-Minkowski space-time. We write the
noncommutative gauge field action as
\begin{equation}
S =  c_1\int F\wedge_\star (*F), \label{NCSg}
\end{equation}
where $*F$ is the noncommutative  dual field-strength tensor. In order to have an
action invariant under the noncommutative gauge transformations
(\ref{ATr}), tensor  $*F$ has to transform covariantly
\begin{equation}
\delta^\star_\alpha (*F) = i [\Lambda_\alpha \ds *F] .\label{*FTr}
\end{equation}
The obvious guess for the noncommutative Hodge dual
\begin{equation}
*F = \frac{1}{2}\epsilon_{\mu\nu\alpha\beta} F^{\alpha\beta}\star
{\rm d}x^\mu\wedge_\star {\rm d}x^\nu \label{*Fobvious}
\end{equation}
does not work since it does not transform covariantly
\begin{equation}
\delta^\star_\alpha (*F)
= {1\over 2}\epsilon_{\mu\nu\alpha\beta} (\delta^\star_\alpha
F^{\alpha\beta})\star {\rm d}x^{\mu}\wedge_\star {\rm d}x^{\nu}
\neq i [\Lambda_\alpha \ds *F] .\nonumber
\end{equation}
Therefore we have to try something else. We
assume that $*F$ has the form
\begin{equation}
*F := {1\over 2}\epsilon_{\mu\nu\alpha\beta}
X^{\alpha\beta}\star {\rm d}x^{\mu}\wedge_\star {\rm d}x^{\nu} ,
\label{*FX}
\end{equation}
where $X^{\alpha\beta}$ are unknown components that should be determined
form the condition (\ref{*FTr}). One way to determine these components is by using the SW map and assuming that
\begin{equation}
X^{\alpha\beta} = F^{0\alpha\beta} + X^{1\alpha\beta} +\dots .
\end{equation}
This would provide us with tensor $X^{\alpha\beta}$ as a function of the commutative field $A_{\mu}^0$.  The other possibility is to make an Ansatz, consistent with (\ref{*FTr}), on the functional dependence of $X^{\alpha\beta}$ on the noncommutative field $A_{\mu}$. When expanded in the deformation parameter, both possibilities should give the same dual field strength. However, the first approach would generate additional ambiguous term in $*F$ coming from the freedom in SW map for $X^{\alpha\beta}$, while in the second approach $*F$ "inherits"  the ambiguities from the SW map for  $A_{\mu}$.  Wishing to keep the ambiguities under control, we choose the second approach. 
Unfortunately, we were unable to find consistent Ansatz for  $X^{\alpha\beta}(A_{\mu})$ in the closed form. Up to the first order in the  deformation parameter we find: 
\begin{eqnarray}
X^{0j} &=& {F}^{0j}-aA_0\star{F}^{0j},\nonumber \\
X^{jk} &=& {F}^{jk}+aA_0\star{F}^{jk}. \label{Xcomp}
\end{eqnarray}
Inserting this into (\ref{*FX}) gives dual field strength that does transform covariantly under the gauge transformations.

Going back to the action (\ref{NCSg}) and writing it more explicitly
we obtain
\begin{equation}
S_g = -\frac{1}{4}\int \Big\{ 2F_{0j}\star e^{-ia\partial_0}X^{0j}
+ F_{ij}\star e^{-2ia\partial_0}X^{ij}\Big\}\star {\rm d}^4 x .
\label{SgComp}
\end{equation}
where the components of $F$ and $X$ are given in (\ref{Fnj}), (\ref{Fij}) and
(\ref{Xcomp}). The terms $e^{-ia\partial_0}X^{0j}$ and
$e^{-2ia\partial_0}X^{ij}$ come from $\star$-commuting basis
1-forms with the components $X^{\mu\nu}$. The constant $c_1$ is fixed in
such a way as to give the good commutative limit of the action
(\ref{SgComp}).

\subsection{Matter field action}

\noindent There are different ways to write a noncommutative gauge invariant action
for spinor matter fields. Since we want to use the cyclicity property
of the integral (\ref{kappaIntCycl}) we have to write the action as an
integral of a maximal form. To this end we introduce the vierbein 1-forms
\begin{equation}
V = V_\mu\star {\rm d}x^\mu = V_\mu^a \gamma_a\star {\rm d}x^\mu ,
\label{VielDef}
\end{equation}
with the Dirac gamma matrices in four dimensions $\gamma_a$ and
$\{\gamma_a, \gamma_b\} = 2\eta_{ab}$. Since we work
in the flat space-time $V_\mu^a = \delta_\mu^a$ and the vierbeins
(\ref{VielDef}) reduce to
\begin{equation}
V = \gamma_\mu\star {\rm d}x^\mu = \gamma_\mu{\rm d}x^\mu.
\label{VielFlat}
\end{equation}
Noncommutative gauge invariance implies the following action
\begin{equation}
S_m = c_2\int \Big( (\overline{D{\psi}})_B \star \psi_A -
\bar{\psi}_B \star (D\psi)_A \Big)\wedge_\star (V\wedge_\star
V\wedge_\star V \gamma_5)_{BA}, \label{SmDef}
\end{equation}
with spinor indices $A, B$ explicitly written (see Appendix
for the explicit calculation) and $\overline{D\psi}
= {\rm d}x^\mu \star \overline{D_\mu^\star\psi}$. Knowing that
under the noncommutative infinitesimal gauge transformations
\begin{equation}
\delta^\star_\alpha V=0, \quad \delta^\star_\alpha \psi
= i\Lambda_\alpha \star\psi, \quad \delta^\star_\alpha \bar{\psi}
= -i \bar{\psi}\star\Lambda_\alpha \nonumber
\end{equation}
one can explicitly show that the action (\ref{SmDef}) is gauge
invariant. In the commutative limit $a\to 0$ this action reduces
to the commutative action for spinor fields \cite{PL1}
\begin{equation}
S_m^0 = \frac{1}{2}{\rm Tr} \int \Big( (D\psi)\bar{\psi} - \psi(\overline{D\psi})\Big)
\wedge (V\wedge
V\wedge V \gamma_5) .\label{SmCommLim}
\end{equation}
In order to write the action (\ref{SmDef}) in a form more convenient for
calculating equations of motion we have to calculate the trace over
spinor indices. We do this calculation explicitly in Appendix and give
the result here:
\begin{eqnarray}
S_{m} &=& c_2\int \Big( (\overline{D{\psi}})_B \star \psi_A -
\bar{\psi}_B \star (D\psi)_A \Big)\wedge_\star (V\wedge_\star
V\wedge_\star V \gamma_5)_{BA} \nonumber\\
&=& 6c_2 \int \Big( i(\overline{D^\star_\mu\psi})\gamma^\mu\star\psi
- \bar{\psi} \star (i\gamma^{\mu}D^\star_{\mu}\psi
\Big) \star {\rm d}^4x
\nonumber
\end{eqnarray}
with
\begin{eqnarray}
&&D^\star_0\psi = \partial^\star_0\psi - iA_0\star\psi,\quad
D^\star_j\psi = \partial^\star_j\psi - iA_j\star e^{-ia\partial_0}\psi,
\nonumber \\
&&\overline{D^*_0\psi} = \partial^\star_0\bar{\psi}
+i\bar{\psi}\star A_0,\quad
\overline{D^\star_j\psi} = \overline{\partial^\star_j\psi}
+ie^{ia\partial_0}(\bar{\psi}\star A_j). \label{BarDDef}
\end{eqnarray}
As we already stressed, 
the twist (\ref{KappaTwist}) leads to the twisted $igl(1,3)$ symmetry.
Since the $igl(1,3)$ algebra contains the conformal subalgebra,
introducing a mass term for the fermions would break the conformal 
symmetry\footnote{We would like to thank R.T. 
Govindarajan for drawing our attention to this point.}
 and therefore the full $igl(1,3)$. Therefore, the twisted $igl(1,3)$ invariant
and the gauge invariant action for the spinor matter field $\psi$ reads:
\begin{equation}
S_m =\frac{1}{2}\int \Big( \bar{\psi} \star i\gamma^{\mu}
(D^\star_{\mu}\psi)
- i{\overline{D^\star_{\mu}\psi}}\gamma^{\mu}
\star\psi \Big) \star {\rm d}^{4}x. \label{SmFinal}
\end{equation}
Notice that we (again) adjusted $c_2$ to get the good commutative limit.

\subsection{Equations of motion}

\noindent Having defined the action in the previous subsections, we are ready to calculate the equations of motion for the fields.
As we are interested only in the first order corrections in
the noncommutativity parameter $a$, we can proceed following two
different procedures. We can vary the complete action $S=S_g+S_m$,
defined by equations (\ref{SgComp}) and (\ref{SmFinal}), with
respect to the noncommutative  fields
and then expand those equation and use the
SW map to obtain the corresponding equations of motion for the
commutative fields with the first order corrections. Alternatively,
we can first expand the complete action  up to the first order in
$a$, use the SW map, and then vary thus obtained
action with respect to the commutative degrees
of freedom. Both procedures give equivalent
equations of motion for the
commutative degrees of freedom with the
first order corrections. Here we present the second option,
writing explicitly the expanded action for the commutative
degrees of freedom and the corresponding equations of motion. 

The expanded action reads
\begin{eqnarray}
S^{\rm expand} &=& S_g^{\rm expand}+S_m^{\rm expand},\nonumber\\
S_g^{\rm expand} &=& -\frac{1}{4}\int {\rm d}^4 x\Big \{ F^0_{\mu\nu}F^{0\mu\nu}
-\frac{1}{2}C^{\rho\sigma}_\lambda x^\lambda F^{0\mu\nu}F^0_{\mu\nu}
F^0_{\rho\sigma}+ \nonumber\\
&&+ 2C^{\rho\sigma}_\lambda x^\lambda F^{0\mu\nu}
F^0_{\mu\rho}F^0_{\nu\sigma}\Big \} , \label{SgExp}\\
S_m^{\rm expand} &=& \frac{1}{2}\int {\rm d}^4 x\Big \{
\bar{\psi}^0 \Big( i\gamma^{\mu}{\cal D}^0_{\mu}\psi^0
+\frac{a}{2} \gamma^jD_0^0 D_j^0\psi^0
+\frac{i}{2}C^{\rho\sigma}_\lambda x^\lambda
\gamma^\mu F_{\rho\mu}^0(D_\sigma\psi^0)\Big) \nonumber\\
&&-\Big( i\overline{D_\mu\psi}^0\gamma^\mu 
-\frac{a}{2}\overline{D_0D_j\psi}^0\gamma^j
+\frac{i}{2}C^{\rho\sigma}_\lambda x^\lambda
\overline{D_\sigma\psi}^0\gamma^\mu F_{\rho\mu}^0\Big)\psi^0
\Big\} .\label{SmExp}
\end{eqnarray}
Note that there are no ambiguous terms in the expanded action coming from the freedom in SW map; all such terms turned out to be total derivative terms and therefore they dropped out from the expanded action. The equations for fermions are:
\begin{eqnarray}
&& i\gamma^{\mu}D^0_{\mu}\psi^0
+\frac{a}{2} \gamma^jD_0^0 D_j^0\psi^0
+\frac{i}{2}C^{\rho\sigma}_\lambda x^\lambda
\gamma^\mu F_{\rho\mu}^0(D_\sigma\psi^0)=0,\nonumber\\
&&-i\overline{D_\mu\psi}^0\gamma^\mu 
+\frac{a}{2}\overline{D_0D_j\psi}^0\gamma^j
-\frac{i}{2}C^{\rho\sigma}_\lambda x^\lambda
\overline{D_\sigma\psi}^0\gamma^\mu F_{\rho\mu}^0=0,\label{ExpEOMPsi}
\end{eqnarray}
while for the gauge field we obtain:
\begin{eqnarray}
&&\partial_\mu F^{0\alpha\mu} + \frac{a}{4}\delta^\alpha_0
F^{0\mu\nu}F^0_{\mu\nu} + 2a F^{0\alpha \mu}F^0_{0\mu}
- C^{\rho\sigma}_\lambda x^\lambda
\Big( \partial_\mu (F_{\rho}^{0\mu}F_{\sigma}^{0\alpha})
+ F^0_{\mu\sigma}(\partial_{\rho}F^{0\mu\alpha} )\Big) \nonumber\\
&& = \bar{\psi}^0\gamma^\alpha \psi^0 + \frac{i}{2}C^{\rho\sigma}_\lambda x^\lambda
\overline{D_\sigma \psi}^0\gamma^\alpha (D_\rho\psi)^0
 + ia \big( \bar{\psi}^0\gamma^\alpha (D_0\psi)^0 -\overline{D_0\psi}^0\gamma^\alpha\psi^0 \big)+\nonumber\\
&&+ \frac{ia}{2}\delta^\alpha_0
\big( \bar{\psi}^0\gamma^0 (D_0\psi)^0
-\overline{D_0\psi}^0\gamma^0\psi^0 \big) .\label{ExpEOMAmu}
\end{eqnarray}
Using either the equations of motions for fermionic fields (\ref{ExpEOMPsi}) or the
expanded action (\ref{SmExp}) we can calculate the
conserved $U(1)$ current up to first order in $a$. Up to total
derivative terms, we obtain:
\begin{eqnarray}\label{cc24}
&&j^0=\bar{\psi^0}\gamma^{0}\psi^0-\frac{a}{2}x^jF^0_{j\sigma}
\bar{\psi^0}\gamma^{\sigma}\psi^0
-\frac{ia}{2}\bar{\psi^0}\gamma^{j}D^0_j\psi^0,\nonumber\\
&&j^k=\bar{\psi^0}\gamma^{k}\psi^0+\frac{a}{2}x^kF^0_{0\sigma}
\bar{\psi^0}\gamma^{\sigma}\psi^0
+\frac{ia}{2}{\overline{D^0_{0}\psi^0}}\gamma^{k}\psi^0 .
\end{eqnarray}

Following the approach of \cite{MajaZadnji} we  discuss deformed
dispersion relations for our fields. Looking at (\ref{SgExp}) we conclude
that there is no modification in the dispersion relation for the photon
field $A^0_\mu$. On the other hand, collecting only the terms quadratic
in the fermionic field from (\ref{SmExp}) we obtain the following equation of motion
\begin{equation}
i\gamma^\mu\partial_\mu \psi + a\gamma^j\partial_0\partial_j \psi
= 0.\label{Disp1}
\end{equation}
Assuming a plane wave solution and 
inserting it in (\ref{Disp1}) leads to
\begin{equation}
(k^0)^2 - (1 - ak_0)^2 \vec{k}^2 =0, \label{Disp2}
\end{equation}
or expanded up to first order in the deforamation parameter $a$
\begin{equation}
(k^0)^2 - \vec{k}^2 + 2ak^0 \vec{k}^2 = 0. \label{Disp2'}
\end{equation}
The dispersion relation is modified in the same
way for all the directions of motion and there is no birefringence
effect. Assuming that $\vec{k} = k\vec{e}_z$ we obtain the following
group velocity
\begin{equation}
v_g = \frac{\partial k_0}{\partial k} = 1- 2ak. \label{GrVelocity} 
\end{equation}
The group velocity\footnote{It is possible to make judicial choice of 
the deformation vector $a^{\mu}$ and thus change the sign on the 
rhs of (\ref{GrVelocity}). This choice gives the group velocity 
bigger than the speed of light (we used natural units, so $c = 1$). 
In view of the latest experimental results from OPERA
collaboration \cite{OPERA} this result could be interesting. 
However, since neutrinos do not couple to the electromagnetic 
field, one would have to construct a non-Abelian gauge theory 
model and check what happens with the dispersion relations in 
that case.} for the massless fermions has momentum  dependence 
in accordance with the previously obtained results 
\cite{nestoACSM}. Note  that this is the first time (to the best 
of our knowledge) that the dispersion follows from the (effective) 
action and  the corresponding equations of motion. 

However, the phenomenological consequences of 
our model  should be taken with care. It is important to remember 
that the corresponding symmetry of the twisted $\kappa$-Minkowski 
space is the twisted $igl(1,3)$ symmetry, and not the 
$\kappa$-Poincar\' e symmetry. 
In our previous analysis \cite{miU1} 
we kept the $\kappa$-Poincar\' e symmetry and that resulted in no 
deformation of the dispersion relations in first order in the 
deformation parameter, see  analysis in \cite{Bolokhov}.  
Also, the $x$-dependent terms in our expanded action clearly 
demand better understanding, possibly in terms of geometric degrees of 
freedom. As in any field 
theory, one needs to understand the renormalization properties 
of the theory before making any predictions. Based on  
the results obtained in field theory on the canonically deformed 
space-time one does expect  additional terms in the action  which 
render theory renormalizable \cite{NCRenorm,MajaZadnji}. Finally, 
the second order corrections in deformation parameter might turn 
out to be essential for deforming of the dispersion relations.

\section{Conclusion and outlook}

In this paper we used the twist formalism to gain a better
understanding of the gauge theory on $\kappa$-Minkowski space-time and to
resolve certain ambiguities we encounter in our previous analysis
\cite{miU1}. The twist formalism provided us with a naturally
defined differential calculus. As a consequence,
we obtained uniquely defined derivatives, thus solving one ambiguity.
Next, in the twist approach the integral has the trace property,
and there is no need to introduce an additional measure
function in the integral. This also means that the limit
of vanishing deformation parameter $a$ reproduces the undeformed
case without the need for additional field redefinitions.
One puzzling feature of the gauge field on $\kappa$-Minkowski
disclosed within the formalism introduced in \cite{kappagft},
was that a gauge field is given in terms of the higher order differential
operator.
This produced "torsion-like"
terms in the field strength which were simply
omitted in the constructed action. In the twist
approach however, the commutation rule of basis
1-forms with functions reproduces the effect of
"higher order differential operator" gauge field without producing
unwanted terms in the action. Furthermore, the
gauge field is enveloping algebra-valued, and one
needs to use the Seiberg-Witten map to express
noncommutative variables (gauge parameter, fields)
in terms of the commutative ones thus keeping the same number of degrees of freedom as
in the commutative case (where the degrees of
freedom are Lie algebra-valued). This mapping introduces
additional ambiguities in the construction of
the effective model. In the previous analysis
we used the additional symmetry requirements to fix
these ambiguities in the constructed action,
while in the twist approach the ambiguities were
completely absent. In a way, this might have
been expected since in the twist approach one
deforms the symmetry of the theory first, and
then consistently applies the
consequences of this deformation on the space-time itself.

We have shown that the twisting of symmetries, as a
way of deforming the algebra of coordinates, is compatible
with the local gauge principle. From this point of view, one could
interpret our (expanded)  model as an example of
non-local and/or non-linear extensions of
electrodynamics  obtained from a more
fundamental theory. The obstruction we have encountered in the construction of Hodge-dual field-strength tensor is a manifestation of the fact that  the
introduction of a noncommutative geometrical structure prevents decoupling of translation
and gauge symmetries, similarly as in general
relativity where translations are part of
"gauge" symmetries (diffeomorphism group). Namely, if one sees   electric charge and magnetic flux as fundamental quantities in electrodynamics, the Maxwell equations can be written in the metric-free form. Only the  relation between the flows of electric charge and magnetic flux, which includes the Hodge dual field strength, introduces the metric degrees of freedom in the
local gauge theory \cite{Hehl}.
Although the mixing of space-time and internal symmetries appeared as a problem in our construction, this is in fact one of the most intriguing property of models based on non-trivial algebras of coordinates.  
One possible way to understand this mixing was offered in the 
framework of Yang-Mills type matrix models \cite{Steinacker}. 
There it was shown that $U(1)$ part of general $U(N)$ gauge group 
can be interpreted as induced gravity coupling to the rest ($SU(N)$) 
gauge degrees of freedom. It would be interesting to see if such an 
interpretation is possible in our framework, by constructing models with larger gauge groups.

\section*{Acknowledgements}
We would like to thank the organisers of the Workshop
on Noncommutative Field Theory and Gravity, Corfu,
September 2010, where some preliminary results were presented.
We would like to thank  Maja Buri\' c, Voja Radovanovi\' c and
Alex Schenkel for fruitful discussion. We gratefully acknowledge
the support of ESF within the framework of the Research Networking
Programme on "Quantum Geometry and Quantum Gravity" and the support of
SEENET-MTP Network through the ICTP-SEENET-MTP grant PRJ-09.  L. J.
acknowledges the support of the Ministry of
Science, Education and Sport of the Republic
of Croatia under the contract 098-0982930-2861. The work of M.D. is supported by
Project No.171031 of the Serbian Ministry of Education and Science.

\appendix

\section{Manipulations with spinors}

\noindent Trace over spinor indices I:
\begin{eqnarray}
S_m &=& \int \Big( (\overline{D\psi})_B \star \psi_A -
\bar{\psi}_B \star (D\psi)_A \Big)\wedge_\star (V\wedge_\star
V\wedge_\star V \gamma_5)_{BA}\nonumber\\
&=& \int \Big( -(\bar{R}^\alpha \psi)_A\star
(\bar{R}_\alpha \overline{D\psi})_B  +
(\bar{R}^\alpha D\psi)_A\star (\bar{R}_\alpha \bar{\psi})_B
\Big)\wedge_\star (V\wedge_\star
V\wedge_\star V \gamma_5)_{BA} \nonumber\\
&=& {\rm Tr} \int \Big( (\bar{R}^\alpha D\psi)\star
(\bar{R}_\alpha \bar{\psi}) -(\bar{R}^\alpha \psi)\star
(\bar{R}_\alpha \overline{D\psi}) \Big)\wedge_\star (V\wedge_\star
V\wedge_\star V \gamma_5) \nonumber
\end{eqnarray}
The minus sign comes from commuting the spinor fields.

\noindent Trace over spinor indices II:
\begin{eqnarray}
S_{m2} &=& - c_2\int \bar{\psi}_B \star (D\psi)_A  \wedge_\star
(V\wedge_\star V\wedge_\star V \gamma_5)_{BA} \nonumber\\
&=& -c_2\int \bar{\psi}_B \star (D^\star_{\mu_1}\psi)_A \star
{\rm d}x^{\mu_1} \wedge_\star
(\gamma_{\mu_2}{\rm d}x^{\mu_2}\wedge_\star \gamma_{\mu_3}{\rm d}x^{\mu_3}
\wedge_\star \gamma_{\mu_4}{\rm d}x^{\mu_4} \gamma_5)_{BA} \nonumber\\
&=& -c_2\int \bar{\psi}_B \star (D^\star_{\mu_1}\psi)_A \star
(\gamma_{\mu_2}\gamma_{\mu_3}\gamma_{\mu_4}\gamma_5)_{BA}
\epsilon^{\mu_1\mu_2\mu_3\mu_4}{\rm d}^4x\nonumber\\
&=& -c_2\int \bar{\psi}_B \star (D^\star_{\mu_1}\psi)_A \star
\epsilon^{\mu_1\mu_2\mu_3\mu_4}i\epsilon_{\mu_2\mu_3\mu_4\mu_5}
(\gamma^{\mu_5})_{BA}{\rm d}^4x\nonumber\\
&=& -6c_2i\int \bar{\psi}_B\gamma^{\mu_1}_{BA} \star (D^\star_{\mu_1}\psi)_A \star
{\rm d}^4x\nonumber\\
&=& -6c_2i\int \bar{\psi}\gamma^{\mu} \star (D^\star_{\mu}\psi) \star
{\rm d}^4x .\nonumber
\end{eqnarray}
In the second line we used:
$$
{\rm d}x^{\mu_1}\wedge{\rm d}x^{\mu_2}\wedge{\rm d}x^{\mu_3}
\wedge {\rm d}x^{\mu_4} =
\epsilon^{\mu_1\mu_2\mu_3\mu_4}{\rm d}^4x, \quad
\epsilon^{\mu_1\mu_2\mu_3\mu_4}\epsilon_{\mu_5\mu_2\mu_3\mu_4}=
-6\delta_{\mu_5}^{\mu_1},$$
and in the third line we used identity valid in four dimensions: $$\gamma_\mu\gamma_\nu\gamma_\rho
= (\eta_{\mu\nu}\eta_{\rho\sigma} -\eta_{\mu\rho}\eta_{\nu\sigma}
+\eta_{\mu\sigma}\eta_{\nu\rho} )\gamma^\sigma
+ i\epsilon_{\mu\nu\rho\sigma}\gamma^\sigma\gamma_5.$$

\end{document}